\documentclass[a4,draftcls,12pt,onecolumn]{IEEEtran}
\usepackage{cite}
\usepackage{epsfig,graphics,mathrsfs,amssymb,amsmath,bm,color,yfonts,txfonts}
\usepackage{subfigure}
\usepackage{algorithm2e}

\begin{document}

\title{Distributed Weighted Sum-Rate Maximization in Multicell MU-MIMO OFDMA Downlink }
\author{Mirza Golam Kibria$^{\dagger}$, Hidekazu Murata$^{\dagger}$ and Jun Zheng$^{\ddagger}$\\
[2mm]
$^{\dagger}$ Graduate School of Informatics, Kyoto University, Kyoto, Japan\\
$^{\ddagger}$ National Mobile Communications Research Laboratory, Southeast University, Nanjing, P.R. China\\
$^{\dagger}$ contact-h25e@hanase.kuee.kyoto.u-ac.jp}

\maketitle

\begin{abstract}
This paper considers distributed linear beamforming in downlink multicell multiuser orthogonal frequency-division multiple access networks. A fast convergent solution maximizing the weighted sum-rate with per base station (BS) transmiting power constraint is formulated. We approximate the nonconvex weighted sum-rate maximization (WSRM) problem with a semidefinite relaxed solvable convex form by means of a series of approximation based on interference alignment (IA) analysis. The WSRM optimization is a two-stage optimization process. In the first stage, the IA conditions are satisfied. In the second stage, the convex approximation of the non-convex WSRM is obtained based on the consequences of IA, and high signal-to-interference-plus-noise ratio assumption. Compared to the conventional iterative distributed algorithms where the BSs exchange additional information at each iteration, the BSs of our proposed solution optimize their beamformers locally without reporting additional information during the iterative procedure.
\end{abstract}

\IEEEpeerreviewmaketitle

\section{Introduction}

The weighted sum-rate maximization (WSRM) is a key element in many network design and optimization methods. However, for a downlink beamforming system, the WSRM problem is known to be NP-hard \cite{Luo}, therefore, very difficult to find the solution. As a result, we have to be reliant on the centralized and computationally very expensive global optimization approaches \cite{Horst, Liu, Joshi} for obtaining the exact solution. However, for a centralized processing based WSRM optimization \cite{Venturino, Wang, Sun}, the overhead for information exchange among the associated base stations (BSs) may be too massive to be implemented in practical systems. Therefore, devising even suboptimal but distributed approaches for WSRM is indeed very important from a practical system design perspective.

There has been a substantial amount of research on suboptimal and distributed WSRM optimization. In \cite{Weeraddana}, the authors proposed a distributed WSRM algorithm based on primal decomposition and subgradient methods, where the original nonconvex WSRM problem is divided into a number of subproblems (one for each base station) and a master problem. In \cite{Choi}, the authors make high signal-to-interference-plus-noise ratio (SINR) approximation to decouple the WSRM problem which involves the beamforming vectors of all BSs into a distributed WSRM problem as a function of local channel state information (CSI), and then solve each decoupled problem by employing a zero-gradient based algorithm. Furthermore, the distributed solutions proposed in \cite{Ka} and \cite{Lindblom} for WSRM are not fully distributed in a sense that at each iteration the BSs have to notify their interference power that depend on other usersf beamformers, and a single user is served per BS in these schemes.  However, all these iterative WSRM optimization designs are for a single career system with the users equipped with a single antenna. The increasing number of antenna elements at the user terminals makes the optimization process even more complex; hence, it is very important to formulate an efficient WSRM solution for multi-antenna users and to evaluate the convergence behavior.

The aim of this study is to propose a distributed WSRM algorithm for the downlink of multicell multiuser multi-input multi-output (MU-MIMO) orthogonal frequency-division multiple access (OFDMA) system. In a multicell scenario, due to the intercell-interference, which affects the respective weighted sum-rate of all the associated BSs, solving the WSRM problem becomes very complicated. We simplify the WSRM problem by decoupling it into multiple distributed problems, each of them solved by the corresponding BS independently. We propose an iterative solution based on a high SINR assumption and the consequences of the interference alignment (IA) technique \cite{Suh, Papai, Gomadam, Peters}. In the iterative procedure, each BS optimizes its own beamformers considering the beamformers used by other BSs as fixed, while keeping the optimization of the weighted sum-rate of the whole system as a global perspective. Unlike \cite{Ka, Lindblom}, our solution does not require the BSs to report the interference powers at each iteration, and therefore, substantially reduces the system overhead.
 
This paper is organized as follows. The multicell MU-MIMO OFDMA system model and the WSRM optimization framework are presented in Section II. In Section III, we address the convex approximation techniques based on IA and a high SINR assumption, and the proposed distributed WSRM solution. In Section IV, we discuss the iterative distributed WSRM algorithm. Section V provides the simulation results and performance analysis. Section VI concludes the paper.
 \\
\textit{Notations:} $(\cdot)^{\rm{H}}$ stands for Hermitian-transpose operation. The Gaussian distribution of complex random variables with mean $\mu$ and variance $\sigma^2$ is defined as $\mathcal{CN}(\mu,\sigma^2)$. Boldface lower-case and upper-case letters define a vector and a matrix, respectively. Operator $\mathrm{diag}(\cdot)$ stacks the diagonal elements of a matrix in a column vector. $\mathbb{C}$ defines a complex space.


\section{System Model and the WSRM Problem Formulation}
In this section, we discuss the system model for the multicell MU-MIMO downlink.  We consider a cellular system of $M$ cells supporting data traffic to $K$ users per cell. We denote the number of BS transmiting antennas and the number of receiving antennas at each user terminal by $N_\mathrm{t}$ and $N_\mathrm{r}$ $(\geq 2)$, respectively. An OFDMA scheme with $N$ subcarriers with 1-cell frequency reuse factor is employed. We also consider non-overlapping subcarrier allocation among the users within a cell. Therefore, the users do not experience intra-cell interference. The subcarrier assignment function $k=f(m,n)$ defines that user $k$ in cell $m$ is assigned with subcarrier $n$. The set of all the BSs is denoted as $\mathcal{M}\triangleq\{1,2,\cdots,M\}$. Thus, the received data vector at user $k$ of cell $m$ over subcarrier $n$, $\bm{y}_{kmn}\in \mathbb{C}^{N_\mathrm{r}\times 1}$ is expressed as
\begin{equation}
{{\bm{y}}_{mkn}}={{\bm{H}}_{mkn}}{{\bm{V}}_{mkn}}{{\bm{s}}_{mkn}}+\sum\limits_{{m}'\in \mathcal{M}\backslash m}{{{\bm{H}}_{{m}'kn}}{{\bm{V}}_{{m}'{k}'n}}{{\bm{s}}_{{m}'{k}'n}}}+{{\bm{z}}_{mkn}}.
\end{equation}
where $\bm{H}_{kmn}\in \mathbb{C}^{{N_\mathrm{r}\times N_\mathrm{t}}} $ is the complex channel matrix between BS $m$ and user $k$, and $\bm{V}_{kmn}\in \mathbb{C}^{{N_\mathrm{t}}\times N_\mathrm{r}} $ denotes the beamformer used by BS $m$ to transmit data to user $k$ on subcarrier $n$. $\bm{s}_{kmn}\sim{\mathcal{CN}(\bm{0},\bm{I}_{N_\mathrm{r}})}$ is data vector transmitted by BS $m$ on subcarrier $n$ that is intended for user $k$.  $\bm{z}_{kmn}\sim{\mathcal{CN}(\bm{0},\bm{I}_{N_\mathrm{r}})}$ denotes the additive white Gaussian noise (AWGN) at user $k$. 
\begin{figure}
  \centering
   \includegraphics[scale=.15]{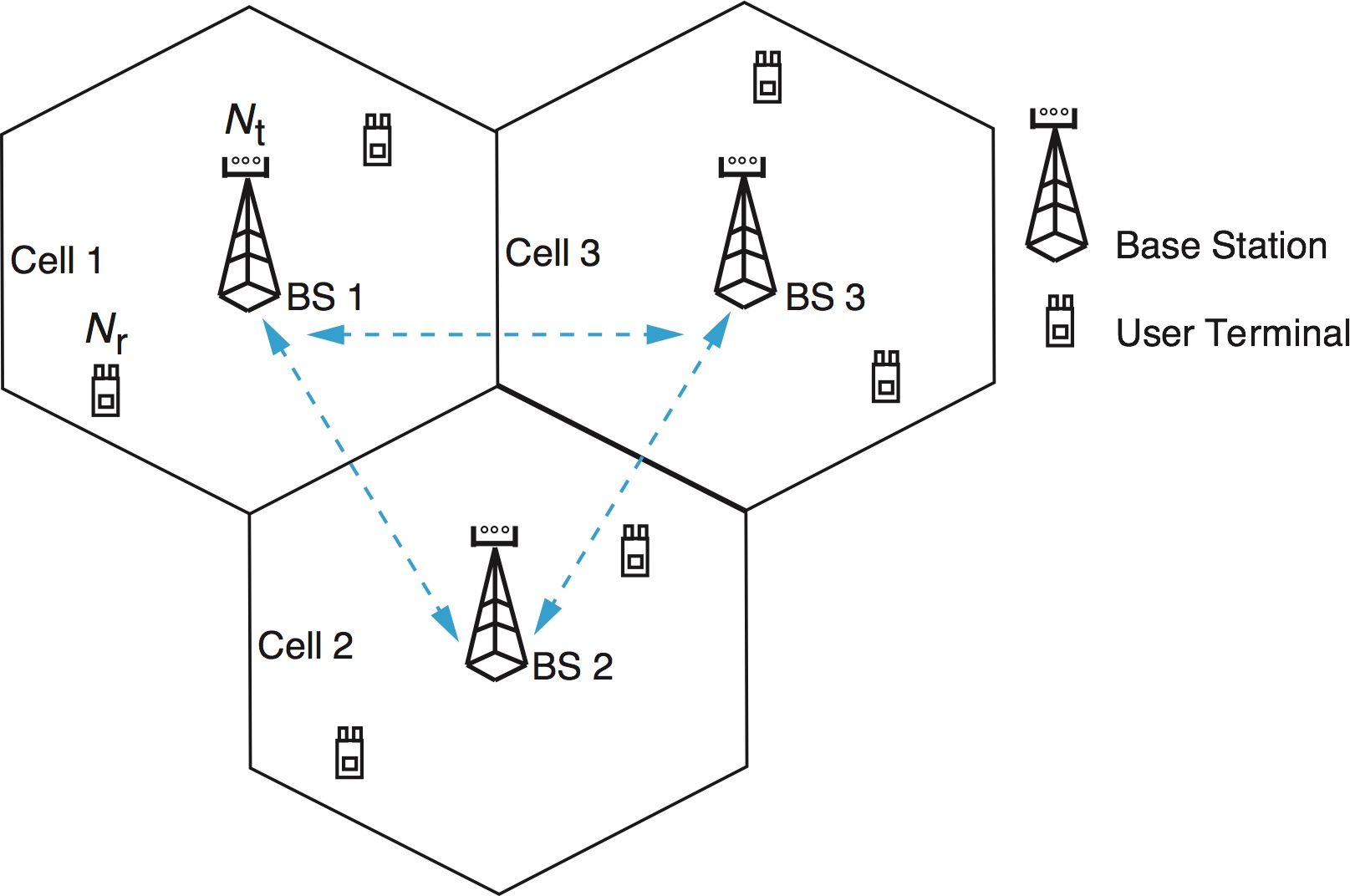}
   \caption{System model for multicell MU-MIMO downlink with distributed processing at each BS. }
   \label{Block}
\end{figure}

The received SINR of user $k$ from cell $m$ scheduled on subcarrier $n$ is given by
\begin{equation}
{\bm{{\gamma }}_{mkn}}=\bm{V}_{mkn}^{\rm{H}}\bm{H}_{mkn}^{\rm{H}}\bm{X}_{mkn}^{-1}{{\bm{H}}_{mkn}}{{\bm{V}}_{mkn}}
\end{equation}
with
$\bm{X}_{mkn}^{-1}={{\bm{I}}}+\sum\limits_{{m}'\in \mathcal{M}\backslash m}{{{\bm{H}}_{{m}'kn}}{{\bm{V}}_{{m}'{k}'n}}\bm{V}_{{m}'{k}'n}^{\rm{H}}\bm{H}_{{m}'kn}^{\rm{H}}}$,
and the corresponding instantaneous downlink rate achieved by user $k$ is formulated as 
\begin{equation}
\label{rate1}
{{R}_{mkn}}=\log \det \left( {{\bm{I}}}+{\bm{{\gamma }}_{mkn}} \right).
\end{equation}
Let us define the set of all the subcarriers scheduled for user $k$ in cell $m$ as ${{\mathcal{S}}_{km}}=\left\{ n|k=f(m,n) \right\}$. Therefore, the total instantaneous rate for user $k$ over all the subcarriers can be expressed by ${{C}_{mk}}=\sum\nolimits_{n\in {{\mathcal{S}}_{km}}}{{{R}_{mkn}}}$. Let $w_{km}$ be the weight associated with user $k$ in cell $m$ that may reflect the quality of the service user $k$ requests in the system or its priority. In this work, the system design objective is to maximize the weighted sum-rate under per BS transmit power constraint. The WSRM problem under BS transmitting power constraints is defined as
\begin{equation}
\label{optzm}
\begin{aligned}
& {\mathop{\max }} \text{~~~~~~~~} \sum\limits_{m\in \mathcal{M}}{\sum\limits_{n\in \mathcal{N}}{{{w}_{mk}}{{R}_{mkn}}}}\\
& \text{subject to}\text{~~}  \sum\limits_{n\in \mathcal{N}}{{\rm{trace}}\left( {{\bm{V}}_{{m}{k}n}}\bm{V}_{{m}{k}n}^{\rm{H}} \right)}\le P_{m,\max }, \hspace{2mm}m=1,...,M,
\end{aligned}
\end{equation}
where $\mathcal{N}\triangleq\{1,2,\cdots,N\}$ is the set of all the subcarriers.
%
%
%
%
%
The rate function in Eq.~\eqref{rate1} is nonconvex in the beamforming matrices ${\bm{V}}_{mkn}$; hence, finding the solution to Eq.~\eqref{optzm} by direct optimization of the beamforming matrices is very hard. As a simplification of this difficulty, we introduce and employ linear receiving filters $\bm{U}_{mkn} \forall m,n $ as auxiliary optimization variables. Now, the received data vector at user $k$, ${{\bm{y}}_{mkn}}$ passes through the linear filter $\bm{U}_{mkn}$, and joint decoding operation is performed to extract the the data vector $\bm{s}_{kmn}$ from the filtered and received vector $\bm{U}_{mkn}^{\rm{H}}{{\bm{y}}_{mkn}}$, which is given by
 \begin{equation}
     \begin{aligned}
     \label{bud1}
	       & \bm{U}_{mkn}^{\rm{H}}{{{{\bm{y}}_{mkn}}}}=\bm{U}_{mkn}^{\rm{H}}{{\bm{H}}_{mkn}}{{\bm{V}}_{mkn}}{{\bm{s}}_{mkn}}+\bm{U}_{mkn}^{\rm{H}}\sum\limits_{{m}'\in M\backslash m}{{{\bm{H}}_{{m}'kn}}}\cdots\\
       & \text{~~~~~~~~~~~~~~~} {\bm{V}}_{m'k'n}{\bm{s}}_{m'k'n}+\bm{U}_{mkn}^{\rm{H}}{{z}_{mkn}}. \\
     \end{aligned}
     \phantom{\hspace{6cm}} 
\end{equation}
%
When both transmitting beamformers and receiving filters are employed together, the rate function in Eq.~\eqref{rate1} can be expressed as
\begin{equation}
     \begin{aligned}
     \label{rate2}
       & {{R}_{mkn}}=\log \det \left( {{\bm{I}}}+\bm{V}_{mkn}^{\rm{H}}\bm{H}_{mkn}^{\rm{H}}{{\bm{U}}_{mkn}}{{\left( \bm{U}_{mkn}^{\rm{H}}\bm{X}_{mkn}{{\bm{U}}_{mkn}} \right)}^{-1}}\right. \cdots \\
       & \text{~~~~~~~~} \left. \bm{U}_{mkn}^{\rm{H}}{{\bm{H}}_{mkn}}{{\bm{V}}_{mkn}}\vphantom{{{\bm{I}}}}\vphantom{{{\bm{I}}}}\right) \\
       & \text{~~~~~}=\log \det \left( {{\bm{I}}}+{{\left( \bm{U}_{mkn}^{\rm{H}}\bm{X}_{mkn}{{\bm{U}}_{mkn}} \right)}^{-1}}\bm{U}_{mkn}^{\rm{H}}{{\bm{H}}_{mkn}}{{\bm{V}}_{mkn}}\right. \cdots \\
       & \text{~~~~~~~~} \left. \bm{V}_{mkn}^{\rm{H}}\bm{H}_{mkn}^{\rm{H}}{{\bm{U}}_{mkn}}\vphantom{{{\bm{I}}}}\right). \\
           \end{aligned}
\end{equation}
It can be immediately justified that there is no capacity loss, i.e., Eq.~\eqref{rate1}=Eq.~\eqref{rate2}, as long as the following optimal receiving filters are applied
\begin{equation}
\label{Uopt}
{{\bm{U}}_{mkn}}=\bm{X}_{mkn}^{-1}{{\bm{H}}_{mkn}}{{\bm{V}}_{mkn}}\hspace{2mm}\forall m,n.
\end{equation}
Therefore, the objective values obtained by solving Eq.~\eqref{optzm} without receiving filters and with receiving filters given in Eq.~\eqref{Uopt} are equal. The advantageous fact of introducing additional optimization variables is that it enables us to perform convex approximation of the nonconvex WSRM problem.  

\section{Convex Approximation based on a high SINR assumption and the consequences of Interference Alignment}

In the convex approximation process, we first make the high SINR approximation of the rate function ${{R}_{mkn}}$ in Eq.~\eqref{rate2} as
\begin{equation}
     \begin{aligned}
     \label{rate3}
       & {{R}_{mkn}}=\log \det \left( {{\bm{I}}}+{{\left( \bm{U}_{mkn}^{\rm{H}}\bm{X}_{mkn}{{\bm{U}}_{mkn}} \right)}^{-1}}\bm{U}_{mkn}^{\rm{H}}{{\bm{H}}_{mkn}}{{\bm{V}}_{mkn}}\right. \cdots \\
       & \text{~~~~~~~~} \left. \bm{V}_{mkn}^{\rm{H}}\bm{H}_{mkn}^{\rm{H}}{{\bm{U}}_{mkn}}\vphantom{{{\bm{I}}}}\right) \\
       & \text{~~~~~}\approx\log \det \left( \bm{U}_{mkn}^{\rm{H}}{{\bm{H}}_{mkn}}{{\bm{V}}_{mkn}}\bm{V}_{mkn}^{\rm{H}}\bm{H}_{mkn}^{\rm{H}}{{\bm{U}}_{mkn}}\right)-\log \det\left(\cdots \right. \\
       & \text{~~~~~~~~}  {{\left. \bm{U}_{mkn}^{\rm{H}}\bm{X}_{mkn}{{\bm{U}}_{mkn}} \right)}}. \\
           \end{aligned}
\end{equation}
In our considered distributed WSRM optimization process, each BS optimizes its own beamforming matrices over all its subcarriers ${{\bm{V}}_{mkn}}$ iteratively considering the beamformers used by other BSs as fixed without exchanging any information during the iterative procedure. Consequently, BS $m$ optimizes its own beamformers with WSRM as the objective function as
\begin{equation}
\label{optzm1}
\begin{aligned}
& {\mathop{\max }} \text{~~~~~~~~~} {\sum\limits_{n\in \mathcal{N}}{{{w}_{mkn}}{{R}_{mkn}}}}\\
& \text{subject to}\text{~~}  \sum\limits_{n\in \mathcal{N}}{{\rm{trace}}\left( {{\bm{V}}_{{m}{k}n}}\bm{V}_{{m}{k}n}^{\rm{H}} \right)}\le P_{m,\max },
\end{aligned}
\end{equation}
where ${w}_{mkn}={w}_{mk}\hspace{1mm}\forall n\in {\mathcal{N}}$. With the high-SINR approximated rate function given in Eq.~\eqref{rate3}, the WSRM problem in Eq.~\eqref{optzm1} can further be equivalently expressed as
\begin{equation}
\label{optzm2}
\begin{aligned}
& \mathrm{max}\hspace{1mm}\sum\limits_{n\in \mathcal{N}} \left( {{w}_{mkn}}\log \det \left( \bm{U}_{mkn}^{\rm{H}}{{\bm{H}}_{mkn}}{{\bm{V}}_{mkn}}\bm{V}_{mkn}^{\rm{H}}\bm{H}_{mkn}^{\rm{H}}{{\bm{U}}_{mkn}}\right)\hspace{0mm}-\hspace{-3mm} \sum\limits_{m'\ne m}\hspace{-1mm}\cdots\right. \\
& \hspace{6mm}\left. w_{m'k'n}\log \det \left(\bm{U}_{m'k'n}^{\rm{H}}{{\bm{H}}_{mk'n}}{{\bm{V}}_{mkn}}\bm{V}_{mkn}^{\rm{H}}\bm{H}_{mk'n}^{\rm{H}}{{\bm{U}}_{m'k'n}}\hspace{-1mm}+\hspace{-1mm}\bm{N}_{m'k'n} \right) \vphantom{\sum\limits_{n\in \mathcal{N}}}\hspace{-1mm} \right) \\
& \mathrm{s.t.} \hspace{3mm}\sum\limits_{n\in \mathcal{N}}{{\rm{trace}}\left( {{\bm{V}}_{{m}{k}n}}\bm{V}_{{m}{k}n}^{\rm{H}} \right)}\le P_{m,\max },\\
\end{aligned}
 \phantom{\hspace{6cm}}
\end{equation}
where ${{\bm{N}}_{m'k'n}}$ is the aggregate leakage interference plus noise at user $k'$ of cell $m'$ scheduled on subcarrier $n$ from the users on the same subcarrier of cells $i\in{\mathcal{M}\backslash m}$, and can be written as
\begin{equation}
{{\bm{N}}_{m'k'n}}=\hspace{-2mm}\sum\limits_{\begin{smallmatrix}
 i\in \mathcal{M}\backslash (m,m') \\
 {j}=f(i, n)
\end{smallmatrix}}{\bm{U}_{m'k'n}^{\rm{H}}{{\bm{H}}_{ik'n}}}{{\bm{V}}_{ijn}}\bm{V}_{ijn}^{\rm{H}}\bm{H}_{ik'n}^{\rm{H}}{{\bm{U}}_{m'k'n}}+{{\bm{U}}_{m'k'n}^{\rm{H}}}{{\bm{U}}_{m'k'n}}.
\end{equation}
We can further rewrite the objective function in Eq.~\eqref{optzm2} as
\begin{equation}
\label{optzmX}
\begin{aligned}
& \mathrm{max}\hspace{1mm}\sum\limits_{n\in \mathcal{N}} \left( {{w}_{mkn}}\log \det \left( \bm{U}_{mkn}^{\rm{H}}{{\bm{H}}_{mkn}}{{\bm{V}}_{mkn}}\bm{V}_{mkn}^{\rm{H}}\bm{H}_{mkn}^{\rm{H}}{{\bm{U}}_{mkn}}\right)\hspace{0mm}-\hspace{-3mm} \sum\limits_{m'\ne m}\hspace{-1mm}\cdots\right. \\
& \hspace{6mm}\left. w_{m'k'n}\log \det \left(\bm{N}_{m'k'n}^{-1}\bm{U}_{m'k'n}^{\rm{H}}{{\bm{H}}_{mk'n}}{{\bm{V}}_{mkn}}\bm{V}_{mkn}^{\rm{H}}\bm{H}_{mk'n}^{\rm{H}}{{\bm{U}}_{m'k'n}}\hspace{0mm}\hspace{0mm} \right)\cdots \right. \\
& \hspace{6mm}-\left. w_{m'k'n}\log \det \left(\bm{N}_{m'k'n}\right)\vphantom{\sum\limits_{n\in \mathcal{N}}}\hspace{0mm} \right). \\
\end{aligned}
 \phantom{\hspace{6cm}}
\end{equation}

Note that the WSRM problem in Eq.~\eqref{optzmX} is still nonconvex since the objective function is nonconvex. Thus, we convexify the objective function based on the implications of the IA process. For any user $k'$ of cell $m'$, the term $\bm{U}_{mkn}^{\rm{H}}{{\bm{H}}_{mkn}}{{\bm{V}}_{mkn}}\bm{V}_{mkn}^{\rm{H}}\bm{H}_{mkn}^{\rm{H}}{{\bm{U}}_{mkn}}$ corresponds to the leakage interference from user $k$ of cell $m$, and $\bm{N}_{m'k'n}$ is the aggregate leakage interference from the users of cells other than cell $m$. When the IA achieved is good enough, the leakage interference from user $k$, $\bm{U}_{mkn}^{\rm{H}}{{\bm{H}}_{mkn}}{{\bm{V}}_{mkn}}\bm{V}_{mkn}^{\rm{H}}\bm{H}_{mkn}^{\rm{H}}{{\bm{U}}_{mkn}}$ lies in the subspace spanned by the interferences from the users of cells $m'\in \mathcal{M}\backslash m$. When we have almost perfect IA achieved, the total leakage interference from all the interfering users, $\bm{U}_{m'k'n}^{\rm{H}}{{\bm{H}}_{mk'n}}{{\bm{V}}_{mkn}}\bm{V}_{mkn}^{\rm{H}}\bm{H}_{mk'n}^{\rm{H}}{{\bm{U}}_{m'k'n}}+\bm{N}_{m'k'n}$ becomes comparable to the background noise at user $k'$ of cell $m'$. Consequently, under a sufficient IA scenario, the largest eigen value of $\bm{N}_{m'k'n}^{-1}\bm{U}_{m'k'n}^{\rm{H}}{{\bm{H}}_{mk'n}}{{\bm{V}}_{mkn}}\bm{V}_{mkn}^{\rm{H}}\bm{H}_{mk'n}^{\rm{H}}{{\bm{U}}_{m'k'n}}$ will be very small. According to \cite{Boyd} and \cite{Yue}, we can approximate $\log \det \left(\bm{N}_{m'k'n}^{-1}\bm{U}_{m'k'n}^{\rm{H}}{{\bm{H}}_{mk'n}}{{\bm{V}}_{mkn}}\bm{V}_{mkn}^{\rm{H}}\bm{H}_{mk'n}^{\rm{H}}{{\bm{U}}_{m'k'n}}+{{\bm{I}}} \right)$ as
\begin{equation}
\begin{aligned}
& \log \det \left(\bm{N}_{m'k'n}^{-1}\bm{U}_{m'k'n}^{\rm{H}}{{\bm{H}}_{mk'n}}{{\bm{V}}_{mkn}}\bm{V}_{mkn}^{\rm{H}}\bm{H}_{mk'n}^{\rm{H}}{{\bm{U}}_{m'k'n}}+{{\bm{I}}} \right)   \\
& \approx \text{trace}\left(\bm{N}_{m'k'n}^{-1}\bm{U}_{m'k'n}^{\rm{H}}{{\bm{H}}_{mk'n}}{{\bm{V}}_{mkn}}\bm{V}_{mkn}^{\rm{H}}\bm{H}_{mk'n}^{\rm{H}}{{\bm{U}}_{m'k'n}}\right).
\end{aligned}
\end{equation}
Consequently, the objective function of the WSRM problem can be reformulated as in expression given below
\begin{equation}
\label{optzm4}
\begin{aligned}
& \mathrm{max}\hspace{1mm} \sum\limits_{n\in \mathcal{N}}\left(  {{w}_{mkn}}\log \det \left( \bm{U}_{mkn}^{\rm{H}}{{\bm{H}}_{mkn}}{{\bm{V}}_{mkn}}\bm{V}_{mkn}^{\rm{H}}\bm{H}_{mkn}^{\rm{H}}{{\bm{U}}_{mkn}} \right)-\hspace{-2mm} \sum\limits_{m'\ne m}\cdots \right. \\
& \hspace{6mm} w_{m'k'n}\left. \text{trace} \left(\bm{N}_{m'k'n}^{-1}\bm{U}_{m'k'n}^{\rm{H}}{{\bm{H}}_{mk'n}}{{\bm{V}}_{mkn}}\bm{V}_{mkn}^{\rm{H}}\bm{H}_{mk'n}^{\rm{H}}{{\bm{U}}_{m'k'n}} \right) \vphantom{\sum\limits_{m'\ne m}}\right).\\
\end{aligned}
 \phantom{\hspace{6cm}}
\end{equation}

 It is advantageous to specify the optimization of the beamforming matrices $\bm{V}_{mkn}$ in terms of their corresponding covariance matrices ${{\bm{W}}_{mkn}}={{\bm{V}}_{mkn}}{{\bm{V}}_{mkn}^{\rm{H}}}$. In order to generate the transmitting symbols with the specified covariances, we can designate the beamforming matrices $\bm{V}_{mkn}$ to be
\begin{equation}
\label{eig}
\bm{V}_{mkn}=\bm{G}_{mkn}\bm{D}_{mkn}^{1/2},
\end{equation}
where $\bm{D}_{mkn}$ is a diagonal matrix and $\bm{G}_{mkn}\bm{D}_{mkn}\bm{G}_{mkn}^{\rm{H}}$ is the eigen-value-decomposition (EVD) of $\bm{W}_{mkn}$. Furthermore, to find the optimal beamformers in terms of covariance matrices, we impose $\bm{W}_{mkn}\succeq O$ to obtain semidefinite program structure of the optimization problem in Eq.~\eqref{optzm4}. As a consequence, $\bm{W}_{mkn}$ becomes the new optimization variables, and we can reformulate Eq.~\eqref{optzm2} as
\begin{equation}
\label{optzmXX}
\begin{aligned}
& \mathrm{max}\hspace{1mm} \sum\limits_{n\in \mathcal{N}}\left(  {{w}_{mkn}}\log \det \left( \bm{U}_{mkn}^{\rm{H}}{{\bm{H}}_{mkn}}{{\bm{W}}_{mkn}}\bm{H}_{mkn}^{\rm{H}}{{\bm{U}}_{mkn}} \right)-\hspace{-2mm} \sum\limits_{m'\ne m}\cdots \right. \\
& \hspace{6mm} w_{m'k'n}\left. \text{trace} \left(\bm{N}_{m'k'n}^{-1}\bm{U}_{m'k'n}^{\rm{H}}{{\bm{H}}_{mk'n}}{{\bm{W}}_{mkn}}\bm{H}_{mk'n}^{\rm{H}}{{\bm{U}}_{m'k'n}} \right) \vphantom{\sum\limits_{m'\ne m}}\right)\\
& \text{subject to.} \hspace{3mm}\text{C1: }\sum\limits_{n\in \mathcal{N}}{{\rm{tr}}\left( {{\bm{W}}_{{m}{k}n}} \right)}\le P_{m,\max }\\
& \hspace{17mm} \text{~C2: }\bm{W}_{mkn}\succeq O \\
& \hspace{17mm} \text{~C3: }\mathrm{rank}\left( {{\bm{W}}_{mkn}}\right)=N_{\rm{r}}. \\
\end{aligned}
 \phantom{\hspace{6cm}}
\end{equation}
In this multi-beam scenario, we consider that the BS transmits $N_{\rm{r}}$ streams to user $k$ in cell $m$. However, the matrix constraint involves NP-Hard difficulty. We drop the rank constraint and obtain an SDP relaxation of Eq.~\eqref{optzmXX}. 
The beamforming matrices $\bm{V}_{mkn}$ are recovered from the covariance matrices according to Eq.~\eqref{eig} by obtaining a rank-$N_{\rm{r}}$ approxmation. To do so, we keep the largest $N_{\rm{r}}$ eigen values while zeroing the rest, and recover $\bm{V}_{mkn}$ as
\begin{equation}
\label{eig1}
\bm{V}_{mkn}=[\bm{v}_1,\bm{v}_2,\cdots, \bm{v}_{N_{\rm{r}}}]\hspace{1mm}\text{diag}(\sqrt{\sigma_1},\sqrt{\sigma_2},\cdots,\sqrt{\sigma_{N_{\rm{r}}}}),
\end{equation}
where $\sigma_i$ is the $i$th largest eigen value of $\bm{W}_{mkn}$ and $\bm{v}_i$ is the associated eigen vector. The intuition is that, after IA is achieved for all the users, the number of interference-free dimensions at receiver $k$ equals $N_{\rm{r}}$. Note that our proposed WSRM solution for multi-antenna users can be straightforwardly formulated for single antenna users; then the optimization variables $\bm{V}_{mkn}$ and $\bm{U}_{mkn}$ become vectors such as $\bm{v}_{mkn}$ and $\bm{u}_{mkn}$, respectively.

\section{Iterative Distributed WSRM algorithm}
In this paper, we propose a convex approximation technique for the nonconvex WSRM optimization problem based on the consequences of IA. This approach iteratively solves the WSRM problem until a convergence point is obtained. The whole optimization process is divided into two independent phases: i) IA phase and ii) post-IA WSRM optimization phase. Each BS performs the optimization process in a distributed manner optimizing its own beamformers while keeping the optimization of the objective function, WSR as a global perspective. For convexification of the WSRM problem, we first make a high SINR assumption, and then subsequently use the implications of the IA process. 

During the IA phase, for obtaining the initial $\bm{V}_{mkn}$ matrices to be used during the WSRM phase, we employ the rank constrained rank minimization (RCRM) technique \cite{Papai}, which reformulates all IA requirements to the requirements involving ranks. Under RCRM approach, the minimization of the sum of the ranks of the interference matrices is performed by minimizing the sum of their corresponding nuclear norms. The rank constraints in RCRM associates the useful signal spaces spanning all available spatial dimensions. This RCRM technique takes a very small number of iterations compared to the max-SINR \cite{Gomadam} and leakage-minimization \cite{Peters} based IA approaches. Note that the IA phase does not aim to maximize the weighted sum-rate, only the IA requirements are fulfilled; hence, complies with the preconditions used for convex approximation. Another important note is that in this paper we do not study the feasibility issue of IA technique. We assume that $N_{\rm{r}}$ degrees of freedom is achievable per user with the IA technique under the considered system model, and use the consequences of the IA technique to facilitate the convex approximation of the nonconvex WSRM problem.

Finally, during the WSRM optimization phase, the IA results are used as the basis. The optimal transmitting beamformers obtained from the RCRM IA phase are used as the initial points in the iterative optimization process. The corresponding receiving beamformers are calculated without incurring any capacity loss that we have already discussed in Section \ref{SM}. Then, we alternatively optimize the transmitting and receiving beamformers until we achieve a convergence point. During the iterative process, there is no inter-BS information exchange. The distributed WSRM algorithm is summarized below
\begin{algorithm}
  \SetAlgoLined
\text{IA Phase}: generates initial $\bm{V}_{mkn} \text{~}\forall m,n$ for the WSRM phase\;
  \textbf{Initialization}: $i=1$, $N_{\rm{iter1}}=10$, $N_{\rm{relz}}=100$\;
  \BlankLine
  \While{$i< N_{\rm{realz}}$}{
   \text{Generate feasible} $\bm{V}_{mkn}, \forall m,n,$ randomly\;\vspace{1mm}
     \text{Run RCRM with} $N_{\rm{iter1}}$ \text{iterations, obtain }$\bm{V}_{mkn}, \forall m,n$\;\vspace{1mm}
     \text{Choose} $\bm{V}_{mkn}, \forall m,n$ \text{that gives the maximum capacity}\;\vspace{1mm}
    }
    \BlankLine
    \text{Post-IA WSRM Phase}\;
    \textbf{Initialization}: $j=1$, $N_{\rm{iter2}}=20$, $\bm{V}_{mkn} \text{~}\forall m,n$ (\text{IA Phase}) \;
    \While{$i< N_{\rm{iter2}}$ {\rm{or}} not converged}{
   \text{Solve Eq.~\eqref{Uopt} and obtain} $\bm{U}_{mkn}, \forall m,n.$\;\vspace{1mm}
     \text{Obtain} $\bm{W}_{mkn} \text{~}\forall m,n$ by solving Eq.~\eqref{optzmXX}\;\vspace{1mm}
     \text{Decompose} $\bm{W}_{mkn}$ as $\bm{G}_{mkn}\bm{D}_{mkn}\bm{G}_{mkn}^{\rm{H}}$, (EVD) \;\vspace{1mm}
      \text{Calculate the optimal} $\bm{V}_{mkn}$ as $\bm{V}_{mkn}=\bm{G}_{mkn}\bm{D}_{mkn}^{1/2}$\;\vspace{1mm}
    }
    \BlankLine
  \caption{WSRM algorithm based on RCRM-IA }
   \label{algo}
\end{algorithm}
\section{Simulation Results and Performance analysis}
In this section, we perform the performance analysis of our proposed convex approximated distributed WSRM solution. We consider a 2-cell system model supporting 2 users each. All the BSs and the users are equipped with 4 antennas and 2 antennas, respectively. The OFDMA scheme with 1-cell frequency reuse factor and 64 subcarriers is considered. Without loss of generality, the user weights are taken as [0.25, 0.54, 0.67, 0.79], which reflect their priorities. The complex coefficients of the channel matrices $\bm{H}_{mkn},\forall m,n$ and $\bm{H}_{m'kn},\forall m'\in{\mathcal{M}\backslash m},n$ are drawn from $\mathcal{CN}(0,1)$. The path-loss and shadowing effects are not considered. To solve the convex approximated problem, we use disciplined convex programming toolbox CVX \cite{CVX} with internal solver SeDuMi \cite{Strum}. As the convergence of the proposed solution strongly depends on the initial of $\bm{V}_{mkn}$s, we follow the IA phase provided in the summarized WSRM algorithm, where we choose the $\bm{V}_{mkn}$ that gives the maximum sum-rate out of $N_{\rm{relz}}$ random initializations.

\begin{figure}[h]
  \centering
   \includegraphics[scale=.14]{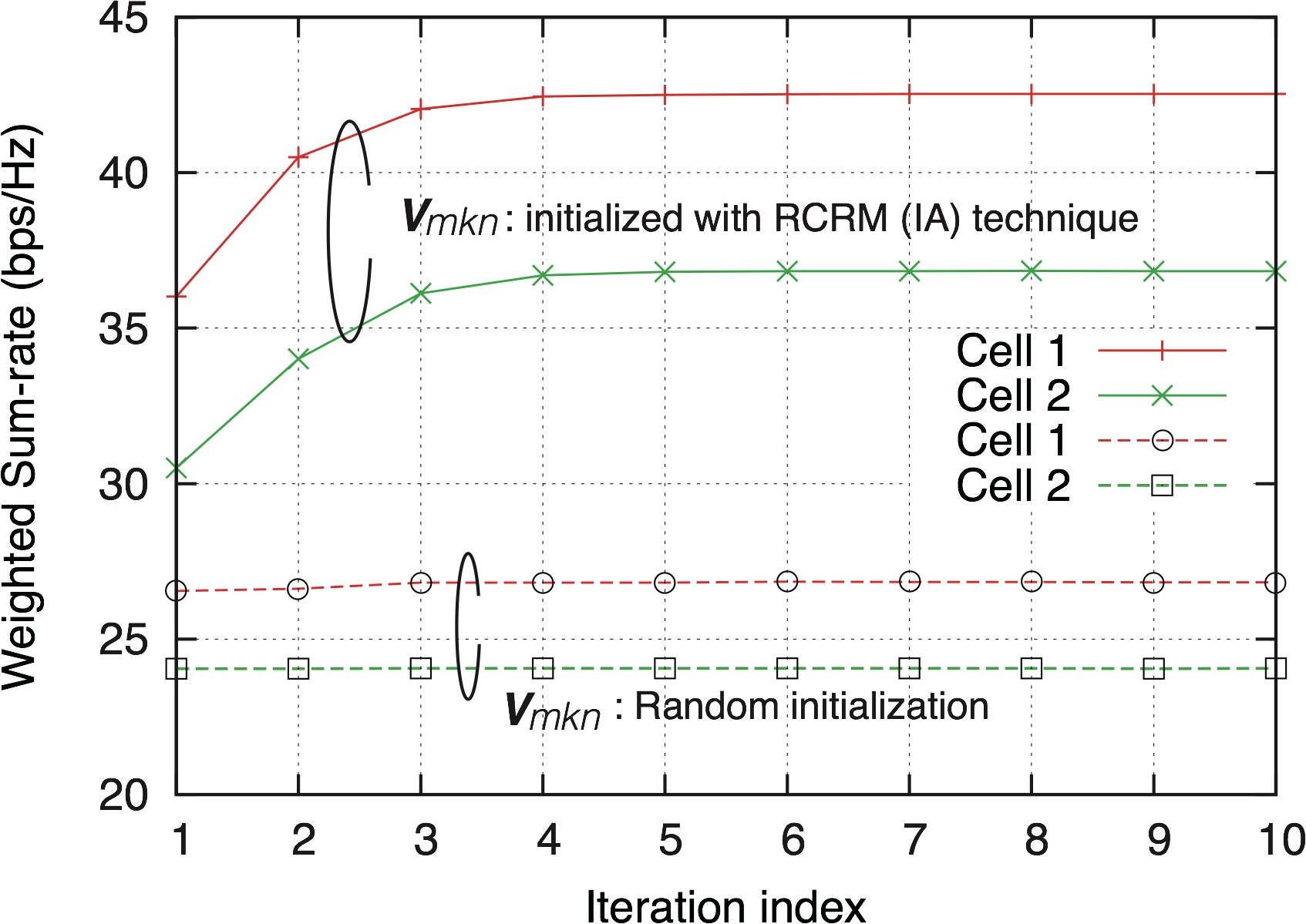}
   \caption{Convergence behaviors comparison}
   \label{WSR}
\end{figure}

We analyze the convergence behavior of our proposed convex approximated WSRM solution in Fig.~\ref{WSR}. We assume that the iterative solution is converged when the difference between two successive iterations is $\le 0.01$. The power budget for each BS is set to 20 dB. We plot the convergence curves for both cells, and compare when initial beamformers $\bm{V}_{mkn}, \forall m,n$ are generated following the IA phase and randomly. We can clearly observe that there exists a significant gap between the IA based initialization and randomly initialized curves. This gain can be regarded as the IA gain. Furthermore, the convergence curve for randomly initialized is not as smooth as the IA based curve. Though the proof of convergence is not provided; however, we have observed that the solution converges all the times for the cases we consider.

\begin{figure}[h]
  \centering
   \includegraphics[scale=.14]{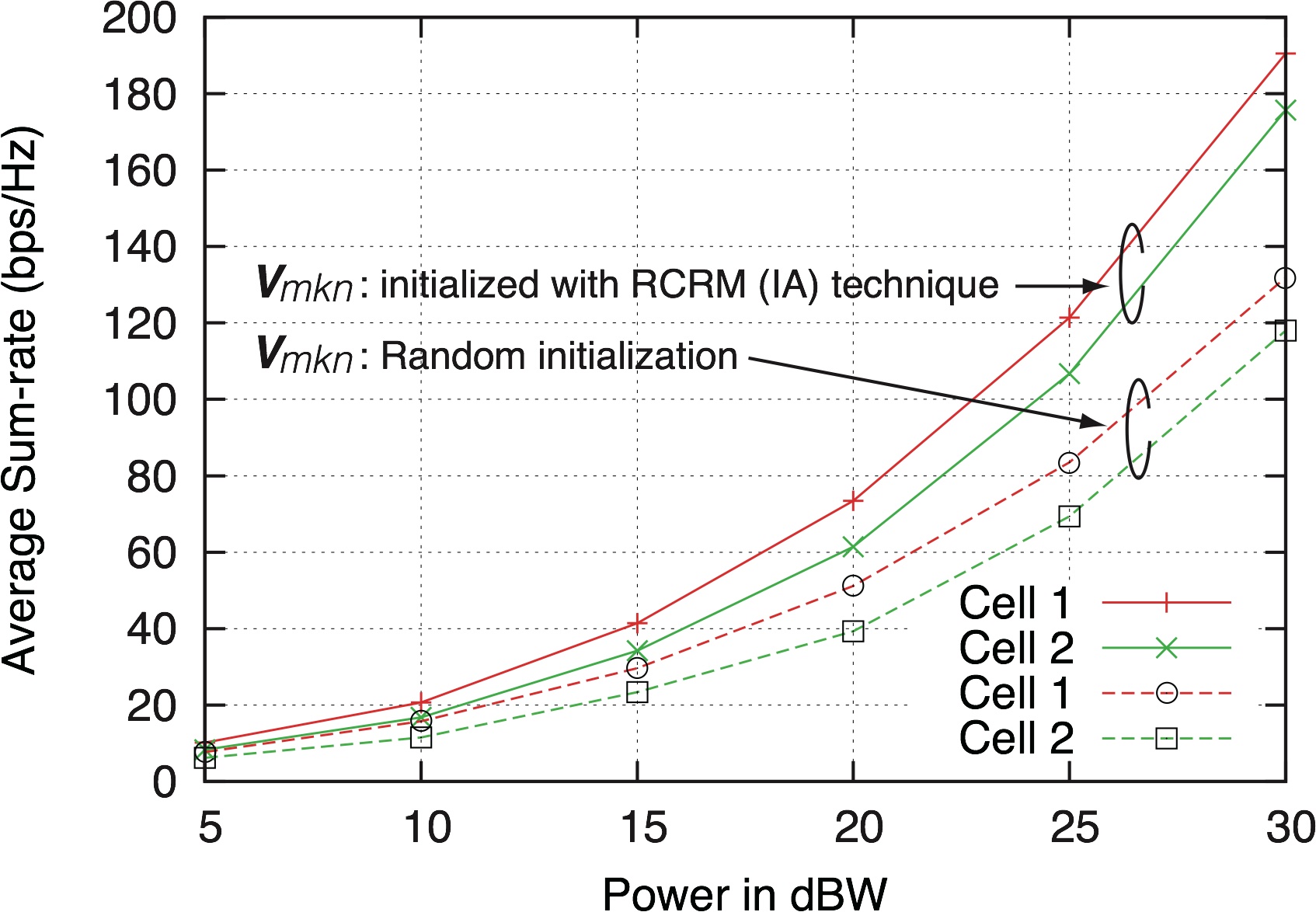}
   \caption{Average sum-rates comparison}
   \label{ASR}
\end{figure}

In Fig.~\ref{ASR}, we evaluate the average sum-rate ($w_{mk}=1,\forall m,k$) performance of our proposed solution. We sweep the BS transmitting power over the range from 5 dBW to 30 dBW. For this experiment, the sum-rates are obtained when the iterative procedures in converged. Like Fig.~\ref{WSR}, we compare the capacities of theIA based and random initialization based convex approximation solutions. We notice that as the BS transmit power increases, the gap between the IA based initialization and the random initialization also increases.  
The rate of capacity increase goes up as the BS transmits with more power.  Like in Fig.~\ref{WSR}, we observe a strong impact of beamformers initialization on the achievable sum-rate.
\vspace{-2mm}
\section{Conclusions}
A distributed approach for WSRM in a multicell MU-MIMO OFDMA is proposed. The proposed algorithm satisfactorily improves the overall system performance with a small amount of base station (BS) cooperations. Each BS optimizes its own beamformers while keeping the whole system WSR as a global perspective. This distributed WSRM technique is indeed favorable in the context of large-size practical communication systems. Unlike other iterative solutions for the WSRM problem, our approach does not require the exchange of information during the iterative optimization operation. Even though the global optimal solution cannot be guaranteed due to the nonconvexity of the original WSRM problem, the numerical results show that our approach requires very small number of iterations for convergence. The proof of convergence of our proposed algorithm has not been studied yet, which is left as our future work.
\section*{Acknowledgement}
This work was supported by the National High Technology Research and Development Program (``863" Program) of China under Grant No. SS2014AA012103, the Open Fund of the State Key Laboratory of Integrated Services Networks, Xidian University, China, under Grant No. ZR2012-01, and the Research Fund of National Mobile Communications Research Laboratory, Southeast University, China, under Grant No. 2014A02.


%
%
%
%
%
%
%
%
%
%
%
%
%
%
%
%
%
%
%
%
%

\end{document}